\documentclass{article}
\usepackage{amsmath}
\usepackage[dvips]{graphicx}
\usepackage{epsfig}
\usepackage{amsmath}
\usepackage{amssymb}
\usepackage{graphics,epstopdf}
\usepackage{amsfonts}
\usepackage{amsthm}
\usepackage[usenames]{color}

\definecolor{AV}{rgb}{0.65,0.0,0}
\definecolor{GC}{rgb}{0,0.0,0.65}
\definecolor{WS}{rgb}{0,0.65,0}

\usepackage[
      colorlinks=true,
      linkcolor=blue,
      urlcolor=blue,
      filecolor=blue,
      citecolor=blue,
      pdfstartview=FitV,
      pdftitle={},
      pdfauthor={},
      pdfsubject={},
      pdfkeywords={},
      pdfpagemode=None,
      bookmarksopen=true
]{hyperref}

\newcommand{\beq}{\begin{equation}}
\newcommand{\eeq}{\end{equation}}
\newcommand{\beqs}{\begin{eqnarray}}
\newcommand{\eeqs}{\end{eqnarray}}
\newcommand{\ra}{\rightarrow}

\newcommand{\beqa}{\begin{eqnarray}}
\newcommand{\eeqa}{\end{eqnarray}}
\newcommand{\beqar}{\begin{eqnarray*}}
\newcommand{\eeqar}{\end{eqnarray*}}

\newcommand{\ee}{\end{equation}}
\newcommand{\eea}{\end{eqnarray}}
\newcommand{\be}{\begin{equation}}
\newcommand{\bea}{\begin{eqnarray}}

\newcommand{\insertplot}[5]{\begin{figure}
 
\hfill\hbox to 0.05in{\vbox to #5in{\vfill
 
\inputplot{#1}{#4}{#5}}\hfill}
 \hfill\vspace{-.1in}
 \caption{#2}\label{#3}
 \end{figure}}
 
\newcommand{\inputplot}[3]{
 \special{ps: 
plotfile #1}
}
\newcounter{fig}

\begin{document}
\title{\bf Charged squashed black holes with negative cosmological constant in odd dimensions}
\author{{\large Yves Brihaye}$^{\dagger}$,
{\large Eugen Radu}$^{\dagger \star}$ 
and {\large Cristian Stelea}$^{\ddagger}$ 
\\ \\
$^{\dagger}${\small Facult\'e des Sciences, Universit\'e de Mons-Hainaut,
B-7000 Mons, Belgium }
\\
$^{\dagger \star}${\small  Institut f\"ur Physik, Universit\"at Oldenburg, Postfach 2503
D-26111 Oldenburg, Germany }
\\
$^{\ddagger}${\small Faculty of Physics, "Alexandru Ioan Cuza"  University, Iasi, Romania}
} 
 
\maketitle

%\ \ \ PACS Numbers: 04.50.+h, 11.10.Kk, 11.15.Kc

\medskip
\begin{abstract}
We construct new electrically charged solutions of the Einstein-Maxwell equations 
with negative cosmological constant in general odd dimensions $d=2n+3 \geq 5$. 
They correspond to higher dimensional generalizations 
of the squashed black hole solutions, 
which were known so far only in five dimensions. 
We explore numerically the general properties of 
such classical solutions and, using a counterterm prescription, 
we compute their conserved charges and discuss their thermodynamics. 
\end{abstract}

%%%%%%%%%%%%%%%%%%%%%%%%%%%%%%%%%%%%%%%%%%%%%%%%%%%%%%%%%%%%%%%%%%%%%%%%%%%%
\section{Introduction}  
%%%%%%%%%%%%%%%%%%%%%%%%%%%%%%%%%%%%%%%%%%%%%%%%%%%%%%%%%%%%%%%%%%%%%%%%%%%%% 
Motivated by recent advances in String/M-Theory (as well as theories with large extra
 dimensions in TeV gravity), black holes in higher
  dimensions have been actively studied in recent years 
with many surprising results
(for reviews of this topic see $e.g.$ \cite{Emparan:2008eg,Obers:2008pj}).
In $d=5$ spacetime dimensions, an interesting class of solutions is provided by 
the so-called squashed Kaluza-Klein (KK) black holes, whose horizon geometry is a squashed three-sphere \cite{Dobiasch:1981vh,Gibbons:1985ac,Rasheed:1995zv,Larsen:1999pp}.
 Their geometry is asymptotic to a non-trivial $S^1$ bundle 
 over a four-dimensional asymptotically flat spacetime, which is 
 also the asymptotic geometry of the Kaluza-Klein magnetic monopole 
 \cite{Sorkin:1983ns,Gross:1983hb}. Such black holes look 
 five-dimensional in the near-horizon region, while at infinity
  (asymptotically) they look like four-dimensional objects with a
   compactified fifth dimension. KK black hole solutions  in the 
   presence of matter fields, are generally found by solving the 
   Einstein equations by brute force. For instance, a solution 
   describing a static KK black hole with a U(1) electric field has been found 
   in \cite{Ishihara:2005dp}, the corresponding non-Abelian generalization
being described in \cite{Brihaye:2006ws}. 
Remarkably, with hindsight, many  such KK solutions can be generated from known
    solutions by applying a `squashing' transformation on suitable geometries \cite{Wang:2006nw,Nakagawa:2008rm,Tomizawa:2008hw,Matsuno:2008fn,Tomizawa:2008rh,Stelea:2008tt}. 

Given the existence of such exotic objects in asymptotically flat settings, it is naturally 
to ask if such solutions do have counterparts in presence of a cosmological constant. 
While on physical grounds one would expect  such objects to exist, so far exact solutions 
have proven to be elusive and one is once again compelled to recourse to numerical methods 
for solving Einstein's field equations. Indeed, finding higher dimensional solutions in 
(anti)- de Sitter ((A)dS) back backgrounds is even harder than the asymptotically flat case, 
as most of the solution-generating techniques break down in presence of a cosmological constant. 
Nonetheless, there are some approximate solutions and also numerical ones that show the existence 
of squashed AdS black holes in five dimensions \cite{Murata:2009jt,Brihaye:2009dm}. Futhermore, extremal charged squashed black holes have been studied in presence of a positive cosmological constant in \cite{Tatsuoka:2011tx}.
  
The main purpose of this work will be to extend some to these results to higher than five dimensions and 
to construct new solutions  of the Einstein-Maxwell theory that describe 
charged squashed AdS black holes in general odd dimensions.  Our construction
is based on the simple observation that one can write the metric of an odd-dimensional 
(round) sphere $S^{2n+1}$ as an $S^1$ fibration over the complex projective space $CP^n$:
\beqs
d\Omega^2_{2n+1}&=&(dz+{\cal A}_n)^2+d\Sigma_n^2,
\label{squashing}
\eeqs
where $d\Sigma_n^2$ is the metric on the `unit' $CP^n$ space, ${\cal A}_n$ 
its Kahler form and $d\Omega_{2n+1}^2$ is the metric on the `unit' round sphere
 $S^{2n+1}$ (we are using the results and conventions from \cite{Hoxha:2000jf}).  
Using this ansatz, the squashing of the sphere $S^{2n+1}$ will be described by a 
squashing parameter $N$ that will multiply the Kahler form ${\cal A}_n$ in (\ref{squashing}). 
Then, in the un-squashed case (here $N=1$), our solutions would correspond to the usual 
Schwarzschild-AdS black holes with spherical horizons, while for $N=0$ they correspond 
to a version of the odd-dimensional AdS black strings found in \cite{Copsey:2006br,Mann:2006yi,Brihaye:2007vm}. 
For general values of $N$ we shall show that there exists a class of black holes in 
AdS-backgrounds whose horizons correspond to a squashed $S^{2n+1}$ sphere.
 Since we were unable to find exact solutions describing such squashed AdS black holes, 
 the configurations in this paper are constructed numerically by matching 
 the near-horizon expansion of the metric to the asymptotic Fefferman-Graham form.

%%%%%%%%%%%%%%%%%%%%%%%%%%%%%%%%%%%%%%%%%%%%%%%%%%%%%%%%%%%%%%%%%%
\section{The model}
%%%%%%%%%%%%%%%%%%%%%%%%%%%%%%%%%%%%%%%%%%%%%%%%%%%%%%%%%%%%%%%%%%
\subsection{The ansatz and the equations of motion}
%%%%%%%%%%%%%%%%%%%%%%%%%%%%%%%%%%%%%%%%%%%%%%%%%%%%%%%%%%%%%%%%%%
We start with the Einstein-Maxwell action principle in $d$-spacetime dimensions:
\begin{eqnarray}
\label{action}
I_0=\frac{1}{16 \pi G_d}\int_{\mathcal{M}} d^d x \sqrt{-g}
 (R-2 \Lambda-F^2)
-\frac{1}{8 \pi G_d}\int_{\partial\mathcal{M}} d^{d-1} 
x\sqrt{-\gamma}K,
\end{eqnarray}
where  $G_d$ is the gravitational constant in $d$ dimensions, 
$\Lambda=-(d-1)(d-2)/(2 \ell^2)$ is the cosmological constant and
 $F=dA$ is the electromagnetic field strength. Here $\mathcal{M}$ 
 is a $d$-dimensional manifold with metric $g_{\mu \nu }$, 
 $K$ is the trace of the extrinsic curvature $K_{ab}=-\gamma_{a}^{c}\nabla_{c}n_{b}$ 
 of the boundary $\partial M$ with unit normal $n^{a}$ and induced metric $\gamma_{ab}$.

By setting the variations of the action (\ref{action}) to zero,
one obtains the Einstein-Maxwell system of field equations:
\beqs
G_{\mu\nu}-\frac{(d-1)(d-2)}{2\ell^2}g_{\mu\nu}=2~T_{\mu\nu},~~~
\nabla_{\mu}F^{\mu\nu}=0,
\label{einstein}
\eeqs
together with a Bianchi identity for the electromagnetic field $dF=0$. Here $G_{\mu\nu}$ is 
the Einstein tensor and $T_{\mu\nu}$ is the stress tensor of the electromagnetic field. 

We consider  here the following parametrization of the $d$-dimensional line element (with odd values of $d=2n+3 \geq 5$):
\begin{eqnarray}
\label{metric} 
ds^2=a(r)\left(dz+N{\cal A}_{n}\right)^2+ \frac{dr^2}{f(r)}+\frac{(d-1)}{(d-4)}r^2d\Sigma^2_{n}-b(r)dt^2~,
\end{eqnarray}
where the $(d{-}3)$--dimensional metric $d\Sigma^2_{n}$ is the metric on the `unit' $CP^{n}$ space, 
${\cal A}_{n}$ is its Kahler form and $N$ is the squashing parameter.
 The direction $z$ is periodic with period $L=\frac{2\pi N\sqrt{(d-1)(d-4)}\ell}{k}$, 
 where $k$ is an integer number. Here the unusual normalization in front of the $CP^n$
  metric $d\Sigma_n^2$ has been chosen in order to make 
 contact in the case $N=0$ to the results obtained previously for the 
 charged black strings in AdS backgrounds \cite{Brihaye:2007vm}, 
 as the normalized metric $\frac{(d-1)}{(d-4)}r^2d\Sigma^2_{n}$ 
 has the same normalization as the unit sphere $S^{d-3}$. 
 
We shall also consider an electric ansatz for the electromagnetic field,
\begin{eqnarray}
\label{U1} 
 A=V(r)dt.
\end{eqnarray}
The Einstein equations with a negative cosmological constant imply then
that the metric functions $a(r)$, $b(r)$ 
and $f(r)$ are solutions of the following equations:
\begin{eqnarray}
\nonumber
\label{ep1} 
f'+\frac{4(d-4)^2 a}{(d-1)^2r^3}N^2
+f\left(\frac{a'}{a}+\frac{b'}{b} \right)
-\frac{2(d-1)r}{\ell^2}
-\frac{2(d-4)(1-f)}{r}
+\frac{4r f}{(d-2)b}V'^2=0,
\end{eqnarray}
\begin{eqnarray}
\label{ep2} 
b''
-\frac{(d-3)(d-4) b}{r^2}(1-\frac{1}{f}) 
+\frac{(d-1)(d-4)b}{ \ell^2 f}
-\frac{(d-3)ba'}{ra}
+\frac{(d-4)b'(1-f)}{rf}
+\frac{(d-1)rb'}{\ell^2f}
\nonumber
\\
-\frac{a'b'}{2a}
-\frac{b'^2}{b}
-\frac{(d-4)^2a\big((d-3)b+2rb'\big)}{(d-1)^2r^4f}N^2
-\left(
\frac{3d-8}{d-2}+\frac{rb'}{(d-2)b}
\right)2V'^2
=0,
\end{eqnarray}
\begin{eqnarray}
\label{ep3} 
\nonumber
a'+\frac{1}{rb'+2(d-3)b}
\bigg(
\frac{2ab}{rf\ell^2}
(
(d-3)(d-4)\ell^2(f-1)-(d-2)(d-1)r^2
)
+2(d-3)ab'
\\
\nonumber
+\frac{2(d-4)^2(d-3)a^2b}{(d-1)^2r^3 f}N^2
+4ra V'^2
\bigg)
=0.
\end{eqnarray}
One can then easily solve the Maxwell equations in (\ref{einstein}) by taking
\beqs
\label{Vr}
V(r)&=&Q(\frac{d-4}{d-1})^{\frac{1}{2}(d-3)}\int^r\frac{1}{r^{d-3}}\sqrt{\frac{b(r)}{a(r)f(r)}}dr+\Phi,
\eeqs
where $Q,~\Phi$ are integration constants, which will be later related to the electric charge
and chemical potential, respectively.
After replacing the corresponding expression of $V'$ in (\ref{ep2}), 
the problem reduced to solving a system of three coupled ordinary differential equations, with given parameters $N$, $\ell$
and $Q$.
 
%%%%%%%%%%%%%%%%%%%%%%%%%%%%%%%%%%%%%%%%%%%%%%%%%%%%%%%%%%%%%%%%%%
\subsection{Asymptotics}
%%%%%%%%%%%%%%%%%%%%%%%%%%%%%%%%%%%%%%%%%%%%%%%%%%%%%%%%%%%%%%%%%%

The corresponding asymptotic expansion of the metric functions for a number $d=2n+3$ of spacetime 
dimension is given by\footnote{Note that in all (odd) spacetime dimensions, one 
finds the asymptotic expression of the Riemann tensor
$R_{\mu \nu}^{~~\lambda \sigma}=-(\delta_\mu^\lambda \delta_\nu^\sigma
-\delta_\mu^\sigma \delta_\nu^\lambda)/\ell^2+O(1/r^{d-4})$.}
\begin{eqnarray}
\nonumber 
a(r)&=&\frac{r^2}{\ell^2}+\sum_{j=0}^{(d-5)/2}a_j(\frac{\ell}{r})^{2j}
+\zeta_a\log(\frac {r}{\ell}) (\frac{\ell}{r})^{d-3}
+c_z(\frac{\ell}{r})^{d-3}+O(\frac{\log r}{r^{d-1}}),
\\
\label{odd-inf}
b(r)&=&\frac{r^2}{\ell^2}+\sum_{j=0}^{(d-5)/2}b_j(\frac{\ell}{r})^{2j}
+\zeta_b\log (\frac {r}{\ell}) (\frac{\ell}{r})^{d-3}
+c_t(\frac{\ell}{r})^{d-3}+O(\frac{\log r}{r^{d-1}}),
\\
\nonumber
f(r)&=&\frac{r^2}{\ell^2}+\sum_{j=0}^{(d-5)/2}f_j(\frac{\ell}{r})^{2j}
+2\zeta_f\log (\frac {r}{\ell}) (\frac{\ell}{r})^{d-3}
+(c_z+c_t+c_0)(\frac{\ell}{r})^{d-3}+O(\frac{\log r}{r^{d-1}}).
\end{eqnarray}   
The  coefficients $a_i$, $b_i$, $f_i$ as well as $\zeta_a$, $\zeta_b$, $\zeta_f$ and $c_0$
depend only on $d,N,\ell$ and have a complicated expression, with no apparent pattern, except for the 
lowest order,
where one finds:
\beqs
a_0&=&\frac{d-4}{d-3}\left(1-\frac{d-4}{d-3}\frac{N^2}{2l^2}\right),~~
~b_0=\frac{d-4}{d-3}\left(1-\frac{(d-4)^2}{(d-3)^2}\frac{N^2}{2l^2}\right),~~{\rm and}~~f_0=b_0+\frac{a_0}{d-2}.
\eeqs

To give a flavour of these expressions, we give here the asymptotic form the solution for $d=7$ only 
(similar expressions were found for $d=9,11$, while the corresponding relations for
$d=5$ were given in \cite{Brihaye:2009dm}) 
\begin{eqnarray}
&&f(r)=\frac{r^2}{\ell^2}+\frac{9}{10}
-\frac{N^2}{5\ell^2}
+\frac{3\ell^2}{320r^2}\left(6-\frac{17N^2}{\ell^2}\right)\left(2-\frac{N^2}{\ell^2}\right)
+\frac{\ell^4}{r^4}\bigg(\frac{9}{1600}+c_t+c_z
\nonumber
\\
&&{~~~~~~}-\frac{3N^2\left(36+\frac{173N^2}{\ell^2}-\frac{94N^4}{\ell^4}\right)}{6400\ell^2}\bigg)-\frac{9\ell^4}{1600r^4}\left(2-\frac{N^2}{\ell^2}\right)\left(6+\frac{25N^2}{\ell^2}-\frac{34N^4}{\ell^4}\right)\log(\frac{r}{\ell})
+O(\frac{\log r}{r^{6}}),
\nonumber
\\
&&a(r)=\frac{r^2}{\ell^2}+\frac{3}{4}-\frac{3N^2}{8\ell^2}
+\frac{3\ell^2}{640r^2}\left(6-\frac{43N^2}{\ell^2}\right)\left(2-\frac{N^2}{\ell^2}\right)
+\frac{\ell^4}{r^4}c_z
\\
\nonumber
&&{~~~~~~~~~~}
-\frac{3\ell^4}{3200r^4}\left(2-\frac{N^2}{\ell^2}\right)\left(18+\frac{97N^2}{\ell^2}-\frac{153N^4}{\ell^4}\right)\log(\frac{r}{\ell})
+O(\frac{\log r}{r^{6}}),
\nonumber
\\
\nonumber
&&b(r)=\frac{r^2}{\ell^2}
+\frac{3}{4}
-\frac{N^2}{8\ell^2}
+\frac{3\ell^2}{640r^2}\left(6-\frac{11N^2}{\ell^2}\right)\left(2-\frac{N^2}{\ell^2}\right)
+\frac{\ell^4}{r^4}c_t
\\
\nonumber
&&
{~~~~~~~~~~~~}
-\frac{3\ell^4}{3200r^4}\left(2-\frac{N^2}{\ell^2}\right)\left(18+\frac{53N^2}{\ell^2}-\frac{51N^4}{\ell^4}\right)\log(\frac{r}{\ell})
+O(\frac{\log r}{r^{6}}).
\nonumber
\end{eqnarray}

For any (odd) value of $d$, terms of higher order in $ \ell/r$ depend on the two 
constants $c_t$ and $c_z$ and also on the charge parameter $Q$ (which enters only the subleading order terms in (\ref{odd-inf})). 
The asymptotic expression of the electric potential is 
$V(r)=\Phi-Q(\frac{d-4}{d-1})^{\frac{1}{2}(d-3)}\frac{\ell Q}{(d-3)r^{d-3}}+\dots$.

The  constants  $(c_t,~c_z)$ are found numerically starting from the following 
expansion of the solutions near the event horizon (taken at constant $r=r_H$) 
and integrating the EM equations towards infinity (this expansion
corresponds to a nonextremal black hole): 
\begin{eqnarray}
\label{eh} 
& a(r)=
a_h+a_1(r-r_H)+O(r-r_H)^2,~
b(r)=b_1(r-r_H)
+O(r-r_H)^2,~
f(r)=f_1(r-r_H)
+O(r-r_H)^2.{~~~}
\end{eqnarray}
The only free parameters in the event horizon expansion are $a_h$, $b_1$.
One finds $e.g.$
\begin{eqnarray}
\label{eh2} 
&&
f_1=\frac{1}{2r_H}
\left(
2(d-4)-\frac{4(\frac{d-4}{d-1})^{d-3}}{a_h(d-2)} \frac{Q^2}{r_H^{2(d-4)}}
+\frac{2(d-1)r_H^2}{\ell^2}
-\frac{4a_h(d-4)^2}{(d-1)^2}\frac{N^2}{r_H^2}
\right),
\\
&&
\nonumber
a_1=\frac{2a_h}{r_H}
\frac{
-\frac{2 (\frac{d-4}{d-1})^{d-3}}{d-2} \frac{Q^2}{r_H^{2(d-4)}}
+\frac{a_h(d-1)r_H^2}{\ell^2}
+\frac{a_h^2(d-4)^2(d-3)}{(d-1)^2}\frac{N^2}{r_H^2}
     }
            {
(d-4)a_h- \frac{2 (\frac{d-4}{d-1})^{d-3}}{d-2} \frac{Q^2}{r_H^{2(d-4)}}
+\frac{a_h(d-1)r_H^2}{\ell^2}
-\frac{2a_h^2(d-4)^2(d-3)}{(d-1)^2}\frac{N^2}{r_H^2}           
             },
\end{eqnarray}
the higher order therms being very complicated,
 and, for simplicity, we shall not list them here. 

Let us also note that in the near horizon region one finds the following expansion of the 
electromagnetic potential,

$
V(r)=V_0+\left( \frac{d-4}{d-1} \right)^{\frac{1}{2}(d-3)}\frac{Q}{r_H^{d-3}} \sqrt{\frac{b_1}{a_h f_1}}(r-r_H)
+O(r-r_H)^2,
$
where $V_0$ is an integration constant. Notice now that one can always set
$V_0=0$ such that $V(r_H)=0$. Then the physical significance of the quantity 
$\Phi$ in (\ref{Vr}) is then that it plays the role of 
the electrostatic potential difference between the infinity and  horizon.

 The condition for a regular (and nonextremal) event horizon is $f'(r_h)>0$, with $b'(r_h)>0$, 
which, for given $r_H$, imposes an upper bound on $Q$,
 \beqs
Q^2<\frac{1}{2}a_h(d-2)\left(\frac{d-1}{d-4}\right)^{d-3}r_H^{2(d-4)}
\left(
d-4+(d-1)\frac{r_H^2}{\ell^2}
-2a_h\frac{(d-4)^2}{(d-1)^2}\frac{N^2}{r_H^2}
\right).
\label{cond}
\eeqs

%%%%%%%%%%%%%%%%%%%%%%%%%%%%%%%%%%%%%%%%%%%%%%%%%%%%%%%%%%%%%%%%%%
\subsection{Global charges and thermodynamical quantities}
%%%%%%%%%%%%%%%%%%%%%%%%%%%%%%%%%%%%%%%%%%%%%%%%%%%%%%%%%%%%%%%%%%
The computation of the mass/energy of these solutions
is basically similar to that performed in  \cite{Mann:2006yi}
for the black string limit of these solutions.
Employing a a counterterm prescription, the total action is
supplemented with 
two boundary terms,
$I=I_0+I_{\mathrm{ct}}^0+I_{\mathrm{ct}}^s,$ 
with $I_{\mathrm{ct}}^0$ the standard AdS
counterterm action 
\cite{Balasubramanian:1999re,Das:2000cu}: 
\begin{eqnarray}
I_{\mathrm{ct}}^0 &=&\frac{1}{8\pi G_d}\int d^{d-1}x\sqrt{-\gamma 
}\left\{ -\frac{d-2}{\ell }-\frac{\ell \mathsf{\Theta }\left( d-4\right) 
}{2(d-3)}\mathsf{R}-\frac{\ell ^{3}\mathsf{\Theta }\left( d-6\right) 
}{2(d-3)^{2}(d-5)}\left(\mathsf{R}_{ab}\mathsf{R}^{ab}-
\frac{d-1}{4(d-2)}\mathsf{R}^{2}\right) 
\right.
\nonumber  
\\
\label{Lagrangianct} 
&&+\frac{\ell ^{5}\mathsf{\Theta }\left( d-8\right) 
}{(d-3)^{3}(d-5)(d-7)}\left( 
\frac{3d-1}{4(d-2)}\mathsf{RR}^{ab}\mathsf{R}_{ab}
-\frac{d^2-1}{16(d-2)^{2}}\mathsf{R}^{3}\right.  \nonumber \\
&&\left. -2\mathsf{R}^{ab}\mathsf{R}^{cd}\mathsf{R}_{acbd}\left. 
-\frac{d-1}{4(d-2)}\nabla _{a}\mathsf{R}\nabla ^{a}\mathsf{R}+\nabla 
^{c}\mathsf{R}^{ab}\nabla _{c}\mathsf{R}_{ab}\right) +...\right\} ,
\end{eqnarray}
where $\mathsf{R}$ and $\mathsf{R}^{ab}$ are the curvature and the 
Ricci tensor associated with the induced metric $\gamma $. 
The series truncates for any fixed dimension, with new terms 
entering at every new even value of $d$, as denoted by the
 step-function ($\mathsf{\Theta }\left( x\right) =1$ provided $x\geq 0$, and vanishes otherwise).

The extraterm $I_{\mathrm{ct}}^{s}$ is due to the presence of
$\log(r/\ell)$ terms in the 
asymptotic expansions of the metric functions  
(with $r$ the radial coordinate) \cite{Skenderis:2000in}:
\begin{eqnarray}
I_{\mathrm{ct}}^{s} &=&\frac{1}{8\pi G_d}\int d^{d-1}x\sqrt{-\gamma 
}\log(\frac{r}{\ell})\left\{  
\mathsf{\delta }_{d,5}\frac{\ell^3 
}{8}(\frac{1}{3}\mathsf{R}^2-\mathsf{R}_{ab}\mathsf{R}^{ab}
)\right.
\nonumber  
\\
&&-\frac{\ell 
^{5}}{128}\left(\mathsf{RR}^{ab}\mathsf{R}_{ab}
-\frac{3}{25}\mathsf{R}^{3} 
-2\mathsf{R}^{ab}\mathsf{R}^{cd}\mathsf{R}_{acbd}\left. 
-\frac{1}{10}\mathsf{R}^{ab}\nabla _{a}\nabla 
_{b}\mathsf{R}+\mathsf{R}^{ab}\Box \mathsf{R}_{ab}
-\frac{1}{10}\mathsf{R}\Box 
\mathsf{R}\right)\delta_{d,7} +\dots
\right\}.\nonumber
\end{eqnarray}

Using these counterterms in the total action,  one computes the boundary stress-tensor: 
\[
T_{ab}=\frac{2}{\sqrt{-\gamma}}\frac{\delta I}{\delta \gamma^{ab}}. 
\]
Following the usual approach, we consider now a standard ADM decomposition of the metric on the boundary,

$
\gamma_{ab}dx^adx^b=-V^2dt^2+\sigma_{ij}(dy^i+V^idt)(dy^j+V^jdt),
$

where $V$ and $V^i$ are the lapse function, respectively the shift vector, and 
$y^i$, $i=1,\dots,d-2$ are the intrinsic coordinates on a closed surface $\Sigma$ 
of constant time $t$ on the boundary. 

Then a conserved charge ${\frak Q}_{\xi }=\oint_{\Sigma }d^{d-2}y\sqrt{\sigma}u^{a}\xi ^{b}T_{ab},$
can be associated with the closed surface $\Sigma$ (with normal $u^{a}$), provided 
the boundary geometry has an isometry generated by a Killing vector $\xi ^{a}$. 

For the line element (\ref{metric}),
the conserved mass/energy $M$ is the charge associated with the time translation 
symmetry, with $\xi =\partial /\partial t$. 
There is also a second charge associated with the compact $z$ direction,
 corresponding to the squashed black hole's gravitational tension ${\mathcal T}$. 

The computation of the boundary stress-tensor
$T_{ab}$ is straightforward and we find the 
following expressions for mass and tension:
\begin{eqnarray}
\label{MT} 
M&=&M_0+M_c^{(d)}~,~{\rm with}~~M_0=\frac{\ell^{d-4}}{16\pi G 
}\big[c_z-(d-2)c_t\big]L{\cal V}_{n}~,
\\
{\mathcal T}&=&{\mathcal T}_0+{\mathcal T}_c^{(d)}~,~{\rm with}~~
{\mathcal T}_0=\frac{\ell^{d-4}}{16\pi G }\big[(d-2)c_z-c_t\big] 
{\cal V}_{n}~,
\end{eqnarray}  
where ${\cal V}_{n}$ is the total area of the `unit' $CP^n$. 
Here $M_c^{(d)}$ and ${\mathcal T}_c^{(d)}$ are  Casimir-like terms,
 \begin{eqnarray}
\nonumber
&&M_c^{(d)}=\frac{\ell^{d-4}{\cal V}_{n}L}{16\pi G}
\bigg[
\frac{1}{12}(1-\frac{N^2}{16\ell^2})^2\delta_{d,5}
-\frac{1}{25600}
( 
1008
+408\frac{N^2}{\ell^2}
-264\frac{N^4}{\ell^4}
+29\frac{N^6}{\ell^6}
)\delta_{d,7}
+\dots\bigg]~,
\\
\nonumber
&&{\mathcal T}_c^{(d)}=\frac{\ell^{d-4}{\cal V}_{n}}{16\pi G}
\bigg[
-\frac{1}{12}
(
1-\frac{19N^2(32\ell^2-7N^2)}{256\ell^4}
)\delta_{d,5}
+\frac{1}{25600}
( 
1008
+9960\frac{N^2}{\ell^2}
-20616\frac{N^4}{\ell^4}
+7667\frac{N^6}{\ell^6}
)\delta_{d,7}
+\dots\bigg]~.
\end{eqnarray}  
The Hawking temperature $T_H$ as obtained from
evaluating the surface gravity or requiring regularity on the Euclidean section is: 
\begin{eqnarray}
T_H=
\frac{1}{4\pi}\sqrt{b_1 f_1},
\label{temp}
\end{eqnarray}
with $f_1$ given by (\ref{eh2}).
The area $A_H$ of the black string horizon is given by 
\begin{eqnarray}
\label{A} 
A_H=r_h^{d-3}{\cal V}_{d-3}L\sqrt{a_h}~.
\end{eqnarray}
As usual, one identifies the entropy of the charged squashed black hole solutions
with one quarter of the event horizon area\footnote{
This relation has been derived in \cite{ Mann:2006yi}
by using Euclidean techniques. This is not  
possible for the electrically charged
solutions in this work, which do not solve the equations of motion for an Euclidean signature. 
However, it can be deduced using the quasi-Euclidean approach.}, $S=A_H/4G_d$.

Finally, for a charged solution, the electric field with respect to a constant 
$r$ hypersurface is given by $E^{\mu}=g^{\mu\rho}F_{\rho\nu}n^{\nu}$. 
The electric charge ${\cal Q}$ of the charged solutions is computed using Gauss' 
law by evaluating the flux of the electric field at infinity:
\begin{eqnarray} 
\label{Qc}  
{\cal Q}=\frac{1}{4\pi G_d}\oint_{\Sigma }d^{d-2}y\sqrt{\sigma}u^{a}n ^{b}F_{ab}=-\frac{Q L{\cal V}_{n}}{4\pi G_d}.
\end{eqnarray} 
If $A_{\mu}$ is the electromagnetic potential, then the electric potential 
$\Phi$, measured at infinity with respect to the horizon 
is defined as:
\beqs
\Phi=A_{\mu}\chi^{\mu}|_{r\ra\infty}-A_{\mu}\chi^{\mu}|_{r=r_h}~,
\eeqs
with $\chi^{\mu}$ a  Killing vector orthogonal to and null on the horizon.
 
%%%%%%%%%%%%%%%%%%%%%%%%%%%%%%%%%%%%%%%%%%%%%%%%%%%%%%%%%%%%%%%%%%
\section{The solutions}
%%%%%%%%%%%%%%%%%%%%%%%%%%%%%%%%%%%%%%%%%%%%%%%%%%%%%%%%%%%%%%%%%%
Solutions of the system (\ref{ep2}), (\ref{ep3})  with the asymptotic behaviour 
 (\ref{odd-inf}), (\ref{eh})
are known to exist in two different limits.
First, for a vanishing squashing parameters, $N=0$,
one recovers the AdS black strings\footnote{Note that these solutions are not known in closed form; however,
they were recovered independently by several different groups, with different
numerical methods.}  discussed in \cite{Copsey:2006br}, \cite{ Mann:2006yi}.
Their generalization with an  electric U(1) field has been considered in \cite{Brihaye:2007vm}.
The second known configuration is found for 
 $N=N_0=\sqrt{\frac{d-1}{d-4}}\ell$, and corresponds to the 
 RNAdS black hole  (with a nonstandard parametrization of the spherical sector).
For our purposes it is helpful to look at this solution from the
viewpoint of the shooting procedure. Assuming that
there is an horizon at $r = r_H$, 
 its expression within the parametrization (\ref{metric})
has $a(r)=\frac{r^2}{\ell^2}$
and
\begin{eqnarray} 
\label{RNAdS}
f(r)=b(r)=
\frac{r^2}{\ell^2}\left(1-(\frac{r_H}{r})^{d-1}\right)
+\frac{d-4}{d-1}\left(1-(\frac{r_H}{r})^{d-3}\right)
+\frac{2(\frac{d-4}{d-1})^{d-3}}{(d-2)(d-3)}
\left(1-(\frac{r}{r_H})^{d-3}\right)
\frac{Q^2\ell^2}{r^{2(d-3)}},
\end{eqnarray}
the global AdS spacetime being recovered for $Q=r_H=0$.

An analytic solution
of the equations (\ref{ep2}) for $N\neq N_0$
 appears to be intractable, and one is contrained to consider numerical methods. 
In our approach, the equations (\ref{ep2}), (\ref{ep3})  
are treated as a boundary value problem for $r \in [r_H,\infty]$ 
(thus we did not consider
the behaviour of the solutions inside the horizon). 
There we employ a collocation
 method for 
boundary-value ordinary
differential equations, equipped with an adaptive mesh selection procedure
\cite{COLSYS}.
The mesh  include $10^3-10^4$ points,
the typical relative accuracy, as estimated by 
the solver being of order $10^{-8}$.
The problem has four input parameter: $r_H$, $N$, $\ell$ and $Q$.
The event horizon data $a(r_H), b'(r_H)$ and the coefficients at infinity $c_t$, $c_z$ and $\Phi$
are read from the numerical output. 

We have found in this way charged squashed black hole solutions
with AdS asymptotics for $d=5,7$ and $9$, the first two cases being studied in a systematic way. 
Thus they are likely to exist for any $d=2n+3\geq 5$. 
For all the solutions 
we studied, the metric functions $a(r)$, $b(r)$, $f(r)$ 
and the electric gauge potential $V(r)$ interpolate
 monotonically between the corresponding values at $r=r_h$ and the 
asymptotic values at infinity, without presenting any local extrema. 

 %%%%%%%%%%%%%%%%%%%%%%%%%%%%%%%%%%%%%%%%%%%%%%%%%%%%%%%%%%%%%%%%%%%%%%%%%% 
\setlength{\unitlength}{1cm}
\begin{picture}(8,6)
\put(-0.5,0.0){\epsfig{file=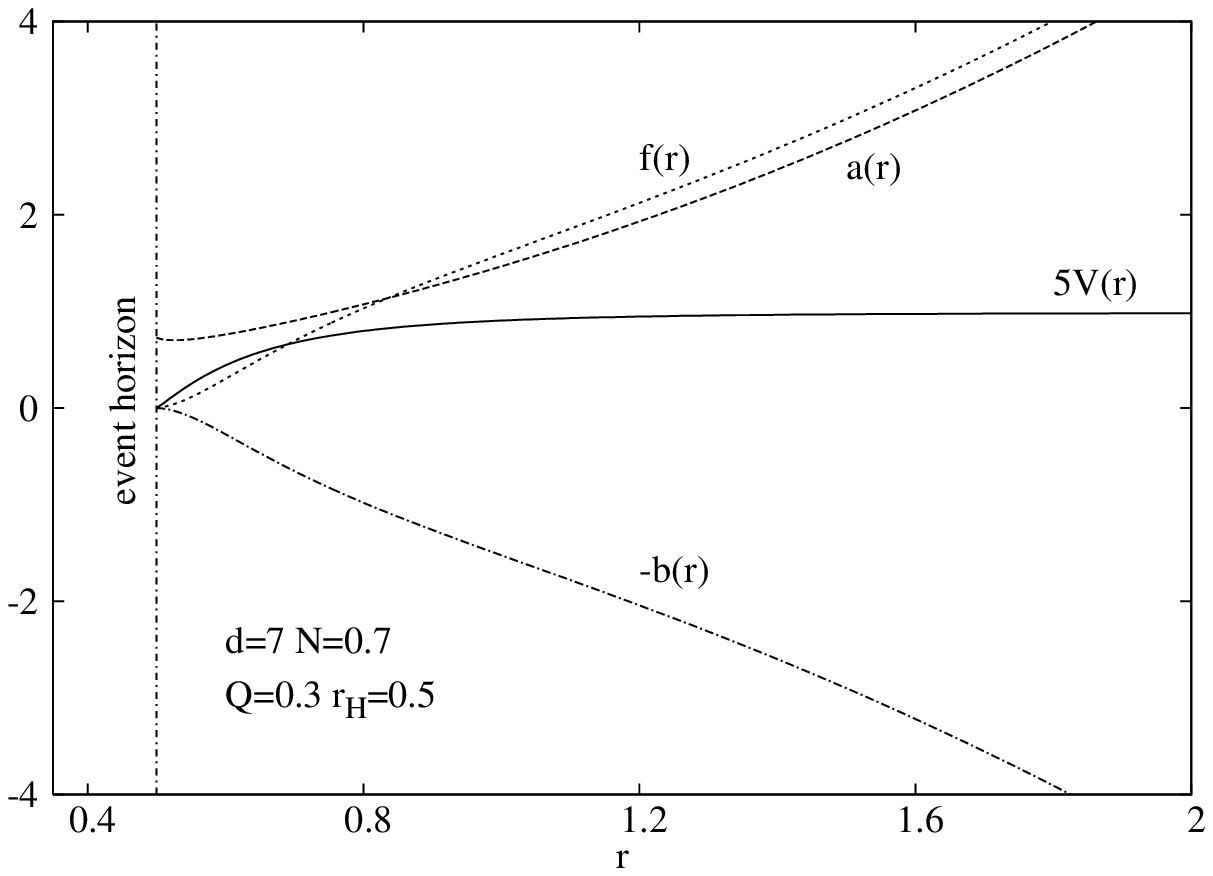,width=8cm}}
\put(8,0.0){\epsfig{file=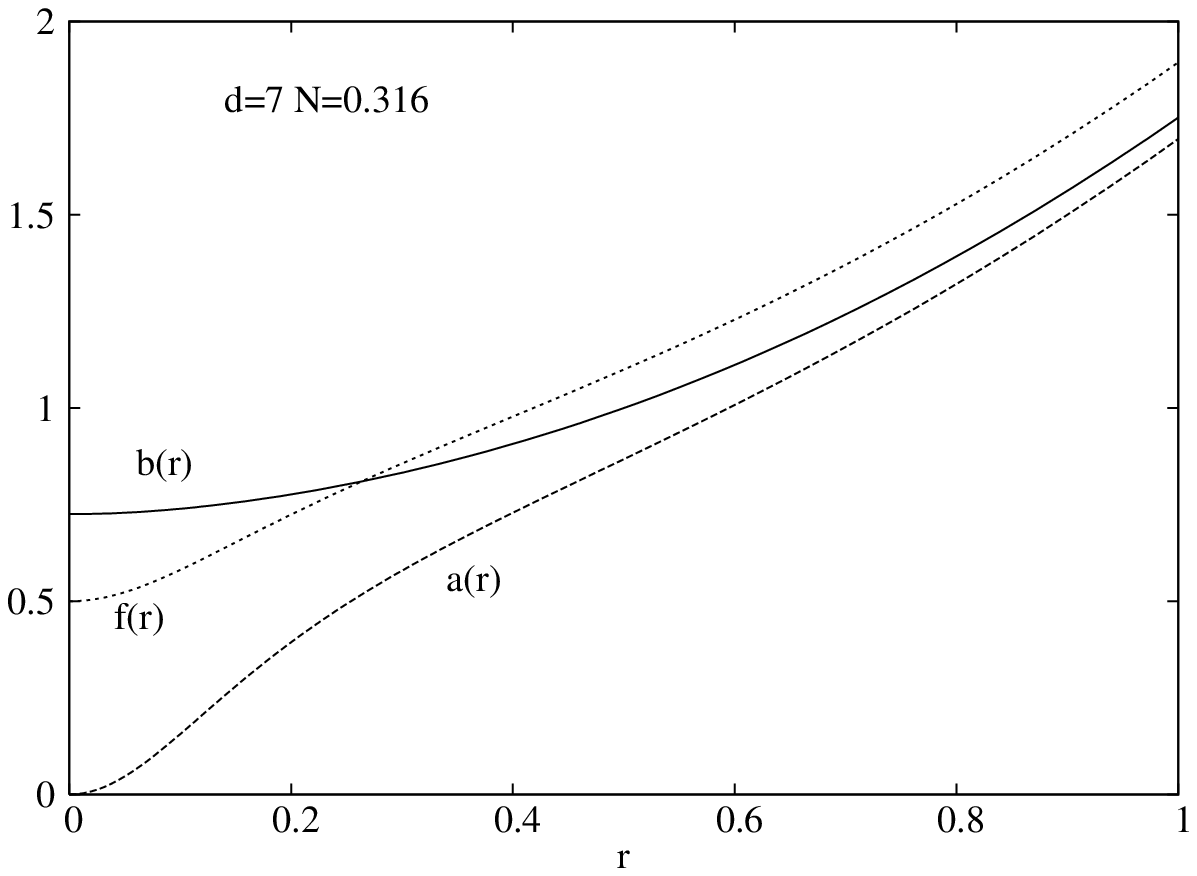,width=8cm}}
\end{picture}
\\
\\
{\small {\bf Figure 1.}  The profile of a typical charged, squashed $d=7$ black hole
(left) is shown together with a that of a $d=7$ soliton (right).
   }
\vspace{0.5cm}
%%%%%%%%%%%%%%%%%%%%%%%%%%%%%%%%%%%%%%%%%%%%%%%%%%%%%%%%%%%%%%%%%%%%%%%%%%%

As a typical example, in Figure $1$ (left) the metric functions $a(r)$, 
$b(r)$ and $f(r)$ as well as the electric gauge potential $V(r)$ 
are shown as functions of the radial coordinate $r$ for a $d=7,~~N=0.7$  
solution with  $r_H=0.5,~Q=0.3,~\ell=1$.
One can see that the term $r^2/\ell^2$ 
starts dominating the profile of  $a(r)$, 
$b(r)$ and $f(r)$ very rapidly, 
which implies a small difference between the metric functions for 
large enough $r$, while the gauge potential $V(r)$ approaches very fast a constant value.

In our numerics, we have fixed 
 the length scale by setting
$\ell=1$ and study the dependence of the solutions
on $r_H$, $N$  and $Q$.
Also, since only $N^2$ appear in the equations, we shall restrict to positive values of $N$.

Starting with the dependence of the solutions on the event horizon radius $r_H$
for fixed squashing and electric charge, our numerical
results
show that the picture valid for the RN-AdS solution is generic. 
For example, for large enough $Q$, one finds 
only one branch of thermally stable black holes, $i.e.$ their entropy increases with the temperature.
Also, for $Q\neq 0$, the solutions exist for a $r_H\geq r_0$, with $r_0$
a solution of the equation
\begin{eqnarray} 
\label{ex3}  
Q^2 N^2-\ell^{2(d-3)}\frac{(d-1)^{d-1}(d-2)}{2(d-4)^{d-2}}
\left(\frac{r_0}{\ell}\right)^{2(d-3)}
\left(
(\frac{r_0}{\ell})^2+\frac{(d-3)(d-4)}{(d-1)^2}
\right)=0.
\end{eqnarray} 
As $r_H\to r_0$, an extremal black hole with $T_H=0$ is approached, the corresponding near horizon solution being
\begin{eqnarray} 
\label{ex1}  
ds^2=v (\frac{dr^2}{r^2}-r^2dt^2)+\frac{(d-1)r_0^2}{(d-4)N^2}\left(dz+N{\cal A}_{n}\right)^2
+ \frac{d-1}{d-4}r_0^2d\Sigma^2_{n},~~
V(r)=V_0+q r,
\end{eqnarray}
where
\begin{eqnarray} 
\label{ex2}  
v=\frac{r_0^2}{(d-2)(d-1)\left(\frac{r_0^2}{\ell^2}+\frac{(d-4)(d-3)^2}{(d-2)(d-1)^2}\right)},~~
q=\frac{v}{\sqrt{2}r_0}\sqrt{(d-2)(d-1)}\sqrt{\frac{r_0^2}{\ell^2}+\frac{(d-3)(d-4)}{(d-1)^2}}.
\end{eqnarray}
In Figure 2, we show the dependence of the event horizon area, mass and tension on the Hawking temperature,
for three different values of $Q$.
These plots retain the basic features 
of the solutions we found for other values of $d,N$ (note that in 
this 
paper we set  $L={\cal V}_{n}/16 \pi G_d=1$ in the numerical values for the mass, 
tension and charge;
also we subtracted the Casimir energy and tension). 
 
 %%%%%%%%%%%%%%%%%%%%%%%%%%%%%%%%%%%%%%%%%%%%%%%%%%%%%%%%%%%%%%%%%%%%%%%%%% 
\setlength{\unitlength}{1cm}
\begin{picture}(8,6)
\put(-0.5,0.0){\epsfig{file=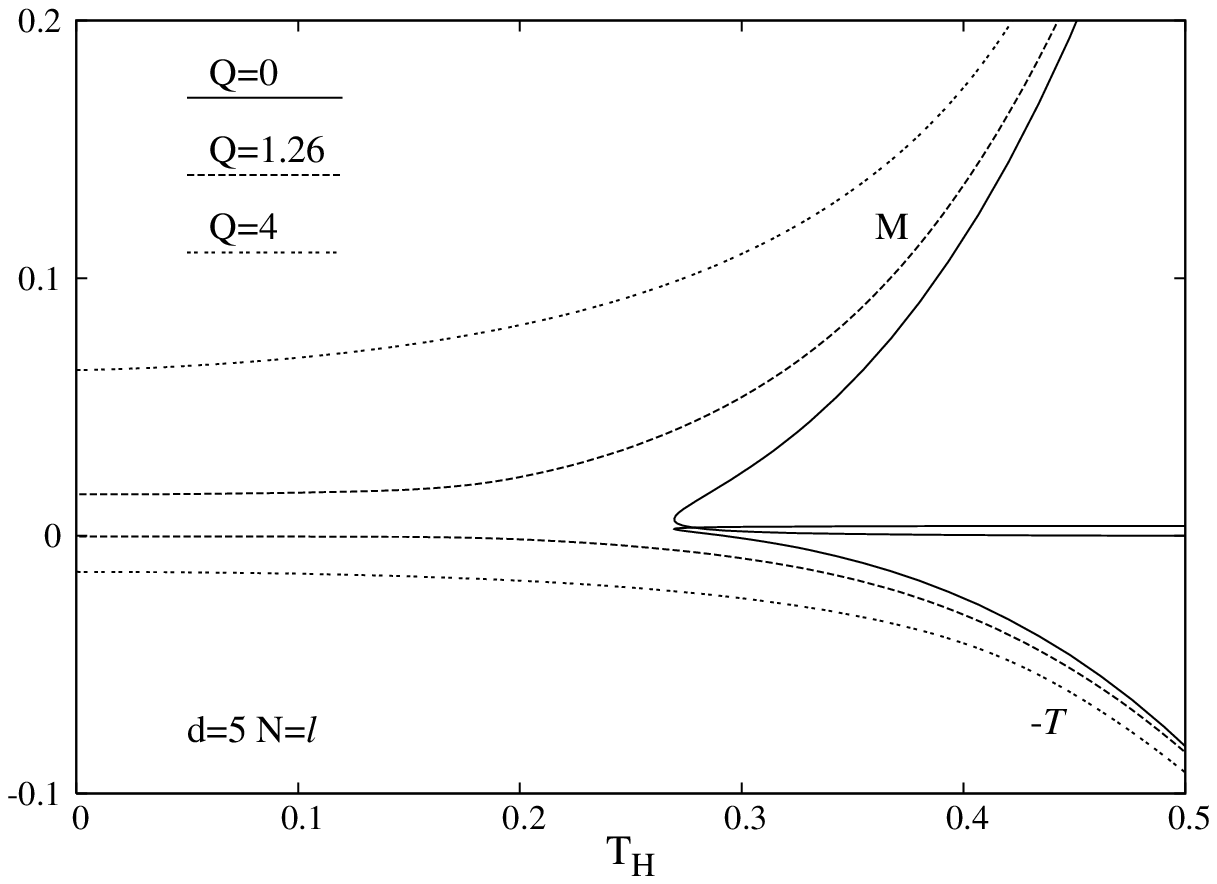,width=8cm}}
\put(8,0.0){\epsfig{file=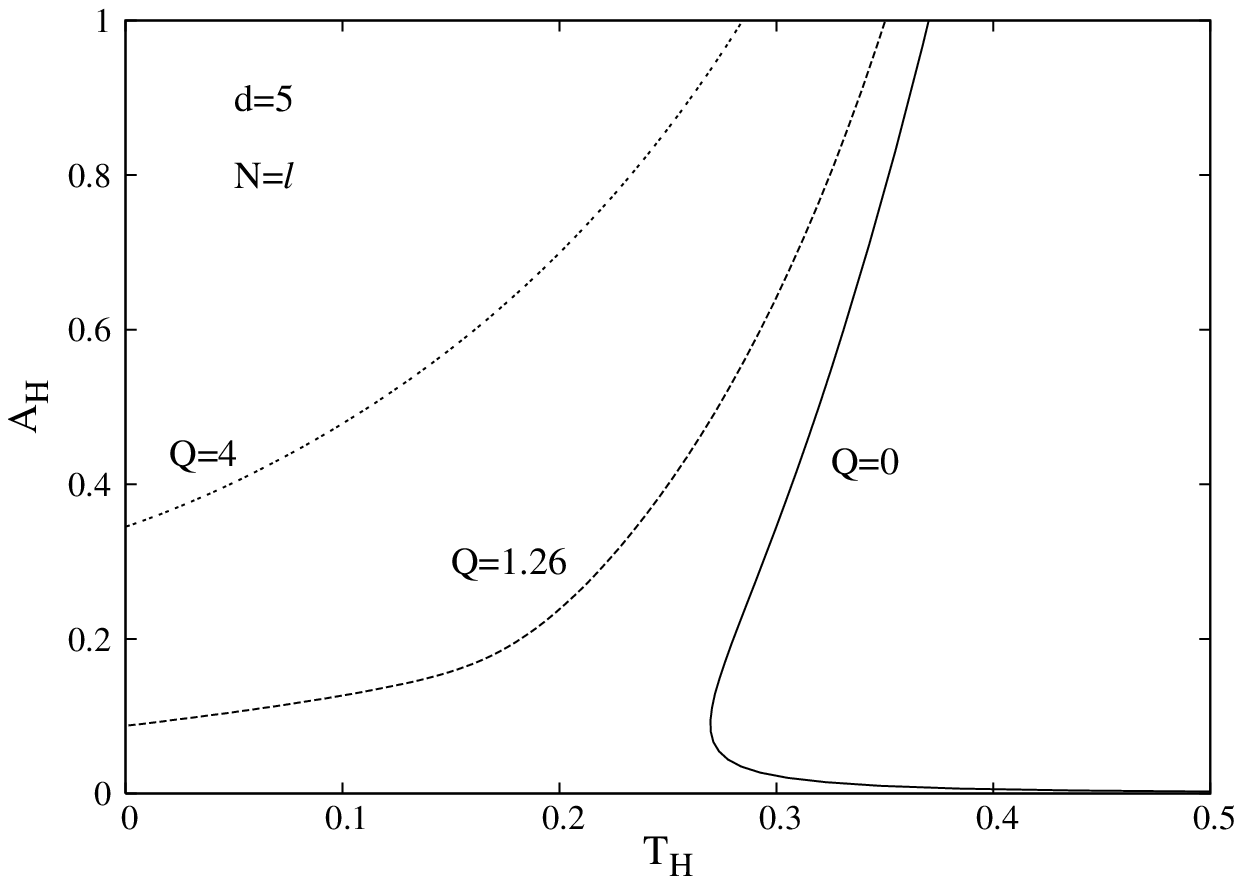,width=8cm}}
\end{picture}
\\
\\
{\small {\bf Figure 2.}  The mass $M$, tension ${\cal T}$ and event horizon area $A_H$
are shown as function of the Hawking temperature $T_H$ for $d=5$
squashed black hole solutions with three different values of the electric charge parameter $Q$.
   }
\vspace{0.5cm}

 %%%%%%%%%%%%%%%%%%%%%%%%%%%%%%%%%%%%%%%%%%%%%%%%%%%%%%%%%%%%%%%%%%%%%%%%%% 
\setlength{\unitlength}{1cm}
\begin{picture}(8,6)
\put(-0.5,0.0){\epsfig{file=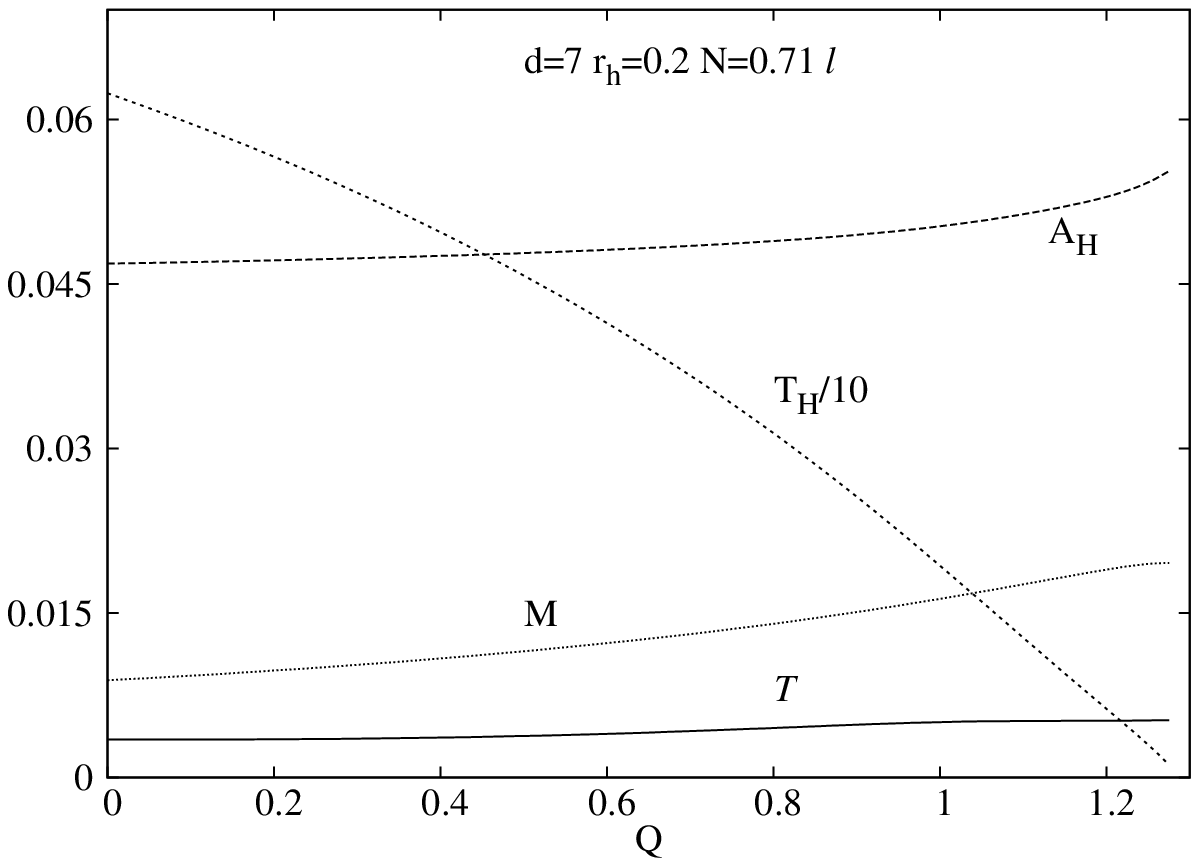,width=8cm}}
\put(8,0.0){\epsfig{file=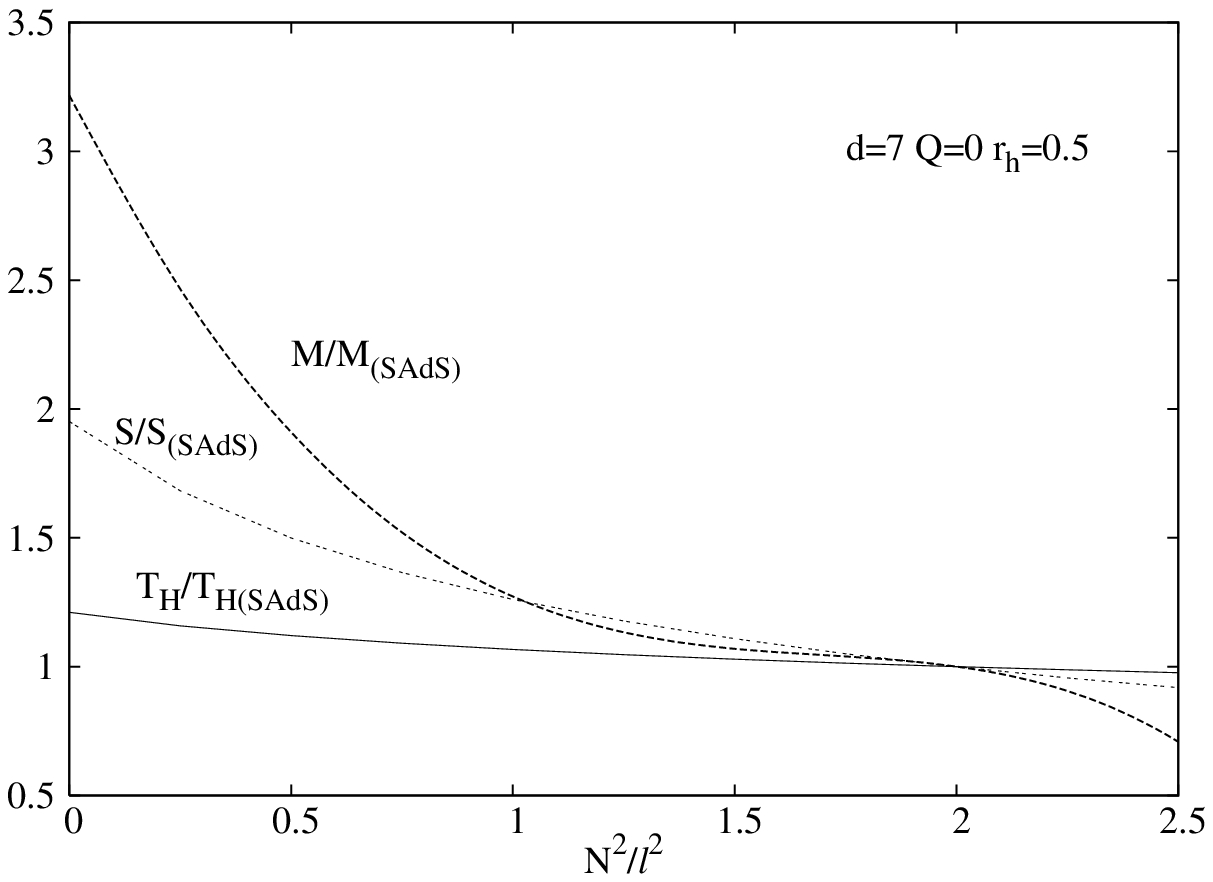,width=8cm}}
\end{picture}
\\
\\
{\small {\bf Figure 3.} A number of  parameters are shown as a function of the electric charge (left)
and  of the squashing parameter $N$,
for  $d=7$ 
squashed black holes with fixed event horizon radius.
   }
\vspace{0.5cm}
%%%%%%%%%%%%%%%%%%%%%%%%%%%%%%%%%%%%%%%%%%%%%%%%%%%%%%%%%%%%%%%%%%%%%%%%%%%
\\
The case $Q=0$ is special, since the solutions exist for any $r_H\geq 0$.
As expected, the properties in this case are rather similar to those of the
Schwarzschild-AdS black hole.
For example, for any $N$, their temperature is bounded
from below, and one finds two branches consisting of smaller (unstable) and large (stable) black
holes.

The $Q=0$ solutions with a vanishing event horizon radius, $r_H=0$, are of special interest, corresponding 
to a deformation of the global AdS background.
These are soliton with a regular origin, the behaviour of the metric functions
as $r\to 0$ being
\begin{eqnarray} 
\label{reg1}  
a(r)=\frac{(d-1)}{(d-4)}\frac{r^2}{N^2}+O(r^4),~~b(r)=b_0+\frac{b_0(d-1)r^2}{(d-4)\ell^2}+O(r^4),
~~
f(r)=\frac{d-4}{d-1}+f_2 r^2+O(r^4),
\end{eqnarray}
in terms of two parameters $b_0,f_2$.
For large $r$, they also approach the asymptotic form (\ref{odd-inf}).

Interestingly, an approximate analytic expression can be constructed in this case, 
viewing the solutions as perturbations around
the global AdS background.
The solutions are constructed order by order, the perturbation parameter
being $\epsilon=N/N_0-1$.
However, the resulting expressions are exceedingly complicated (and not
very enlightening)
 except for the first order in $\epsilon$.
Then one finds the following distorsion of the global
AdS background
\begin{eqnarray} 
\label{pert1}
&&
f(r)=\frac{r^2}{\ell^2}+\frac{1}{4}
+\epsilon \left (-\frac{5}{6}+\frac{\ell^2}{4r^2}
\left(
-1+4(\frac{1}{3}+\frac{\ell^2}{16r^2})\log(1+\frac{4r^2}{\ell^2})
\right )
\right),
\\
\nonumber
&&
a(r)=\frac{r^2}{\ell^2}
+\epsilon 
\left(-1+\frac{\ell^2}{4 r^2}\log(1+\frac{4r^2}{\ell^2})
 \right),
 ~~b(r)=\frac{r^2}{\ell^2}
+\epsilon 
\left(-\frac{1}{2}+\frac{\ell^2}{12r^2}\log(1+\frac{4r^2}{\ell^2})
 \right),
\end{eqnarray}
for $d=5$,
\begin{eqnarray} 
\label{pert2}
&&
f(r)=\frac{r^2}{\ell^2}+\frac{1}{2}
+\epsilon \left (-\frac{4}{5}+\frac{21\ell^2}{20r^2}
\left(
1+\frac{5\ell^2}{7 r^2}
-\frac{6\ell^2}{7r^2}(1+\frac{5\ell^2}{12 r^2})
\log(1+\frac{2r^2}{\ell^2})
\right )
\right),
\\
\nonumber
&&
a(r)=\frac{r^2}{\ell^2}
+\epsilon 
\left(-\frac{3}{2}+\frac{3\ell^2}{2 r^2}\left(1-\frac{\ell^2}{2r^2}\log(1+\frac{2r^2}{\ell^2})\right)
 \right),~
b(r)=\frac{r^2}{\ell^2}+\frac{1}{2}
+\epsilon 
\left(-\frac{1}{2}+\frac{3\ell^2}{10r^2}\left(1-\frac{\ell^2}{2r^2}\log(1+\frac{2r^2}{\ell^2})\right)
 \right),
\end{eqnarray}
for $d=7$, its expression for higher $d$ becoming too complicated to display it here.
The nonperturbative solutions with arbitrary $N$ are constructed again numerically,
a typical soliton profile being exhibited in Figure 1 (right).

Returning to solutions with an event horizon, we show in Figure 3 the dependence of the
$d=7$ squashed black holes on the electric charge  (left) 
and on the squashing parameter $N$ (right), for a fixed event horizon radius.
One can see for example that, for  given $r_H$, the mass,
tension and entropy of solutions decrease 
monotonically with $N$.
Nontrivial solutions appear to exist for all values of $N$,
the value  $N_0=\sqrt{\frac{d-1}{d-4}}\ell$ separating prolate metrics from the
oblate case ($N>N_0$).

%%%%%%%%%%%%%%%%%%%%%%%%%%%%%%%%%%%%%%%%%%%%%%%%%%%%%%%%%%%%%%%%%%%%%%%%%%%%
\section{Further remarks }
%%%%%%%%%%%%%%%%%%%%%%%%%%%%%%%%%%%%%%%%%%%%%%%%%%%%%%%%%%%%%%%%%%%%%%%%%%%%% 
 The specific class of solutions that we constructed here are likely to be relevant
 in connection to the so-called `fragility' of the AdS  black holes. 
 As described in \cite{McInnes:2010ti} (see also \cite{McInnes:2011eg}) 
 it turns out that AdS black holes can become unstable to stringy effects when their horizon 
 geometries are sufficiently distorted. 
 By the AdS/CFT correspondence, this implies that 
 their field theory on the boundary can also become unstable. 
 In five dimensions this phenomenon was explicitly checked in 
 \cite{Murata:2009jt,Brihaye:2009dm}. More specifically, if the spherical
horizon of a black hole is sufficiently squashed then the theory on the boundary ceases 
to be stable since the scalar sector of the theory becomes tachyonic for 
sufficiently large squashing. At first sight, the squashed AdS black hole
geometry does not exhibit any signs of gravitational and/or thermodynamic pathologies 
for this same amount of squashing of the black hole horizon that causes the 
instability of the boundary field theory. However, as shown in \cite{McInnes:2010ti}, 
the source of the instability in the background of the AdS squashed black hole is 
related to the stringy pair-production of branes in this geometry, a phenomenon discovered
 by Seiberg and Witten \cite{Seiberg:1999xz}.

 The $Q=0$ configurations we described in this paper reduce in five dimensions
   to the squashed black holes studied in \cite{Murata:2009jt,Brihaye:2009dm}, 
   while providing the needed generalizations to higher than five odd dimensions. 
   As such, they will be directed relevant in connection to the fragility of the squashed AdS black holes 
   in higher than five dimensions. 
   In particular, we expect that for certain values of the squashing parameter
    the squashed geometry becomes unstable to the phenomenon of pair-production 
    of branes and this will signal potential instabilities of the field theory on the boundary.
     Let us consider as an example the seven-dimensional squashed geometry.
      In this case the Ricci curvature scalar of the rescaled boundary geometry has the expression:
\beqs
R_7&=&\frac{12\ell^2-N^2}{\ell^4}.
\eeqs 
It is now clear that for sufficiently large values of 
$N$ the Ricci scalar curvature can be constant and negative. 
This is a sufficient condition for the onset of the non-perturbative 
Seiberg-Witten stringy effect of pair-production of branes in this geometry. 
According to the AdS/CFT conjecture, this signals an instability in the boundary conformal field theory. 
 
As avenues for further research, let us mention that we have also found solutions that 
correspond to black hole geometries with squashed horizons in absence of the cosmological constant.
These solutions 
have a different asymptotics expansion than (\ref{odd-inf}) and will be discussed elsewhere.
 Also, one should be able to further generalize these solutions to include the effects of rotation. 
 However, as these new solutions deserve separate studies, we shall report them in further work. 

%%%%%%%%%%%%%%%%%%%%%%%%%%%%%%%%%%%%%%%%%%%%%%%%%%%%%%%%%%%%%%%%%%%%%%%%%
\bigskip
\noindent
{\bf\large Acknowledgements} \\
Y.B. is grateful to the
Belgian FNRS for financial support.
E.R. gratefully acknowledges support by the DFG. 
The work of C.S. was financially supported by POSDRU through the POSDRU/89/1.5/S/49944 project.
%%%%%%%%%%%%%%%%%%%%%%%%%%%%%%%%%%%%%%%%%%%%%%%%%%%%%%%%%%%%%%%%%%%%%%%%%
 
%%%%%%%%%%%%%%%%%%%%%%%%%%%%%%%%%%%%%%%%%%%%%%%%%%%%%%%%%%%%%%%%%%%%%%%%%%%%%%%%%%%

%%%%%%%%%%%%%%%%%%%%%%%%%%%%%%%%%%%%%%%%%%%%%%%%%%%%%%%%%%%%%%%%%%%%%%%%%%%%%%

\begin{thebibliography}{99}
%%%%%%%%%%%%%%%%%%%%%%%%%%%%%%%%%%%%%%%%%%%%%%
%\cite{Emparan:2008eg}
\bibitem{Emparan:2008eg}
  R.~Emparan and H.~S.~Reall,
  %``Black Holes in Higher Dimensions,''
  Living Rev.\ Rel.\  {\bf 11}, 6 (2008)
  [arXiv:0801.3471 [hep-th]].
  %%CITATION = 00222,11,6;%%
%%%%%%%%%%%%%%%%%%%%%%%%%%%%%%%%%%%%%%%%%%%%%%
%\cite{Obers:2008pj}
\bibitem{Obers:2008pj}
  N.~A.~Obers,
  %``Black Holes in Higher-Dimensional Gravity,''
  Lect.\ Notes Phys.\  {\bf 769}, 211 (2009)
  [arXiv:0802.0519 [hep-th]].
  %%CITATION = LNPHA,769,211;%%  
 %%%%%%%%%%%%%%%%%%%%%%%%%%%%%%%%%%%%%%%%%%%%%% 
 %\cite{Dobiasch:1981vh}
\bibitem{Dobiasch:1981vh}
  P.~Dobiasch and D.~Maison,
  % ``Stationary, Spherically Symmetric Solutions Of Jordan's Unified Theory Of
  %Gravity And Electromagnetism,''
  Gen.\ Rel.\ Grav.\  {\bf 14}, 231 (1982).
  %%CITATION = GRGVA,14,231;%%
%%%%%%%%%%%%%%%%%%%%%%%%%%%%%%%%%%%%%%%%%%%%%%
%\cite{Gibbons:1985ac}
\bibitem{Gibbons:1985ac}
  G.~W.~Gibbons and D.~L.~Wiltshire,
  %``Black Holes In Kaluza-Klein Theory,''
  Annals Phys.\  {\bf 167}, 201 (1986)
  [Erratum-ibid.\  {\bf 176}, 393 (1987)].
  %%CITATION = APNYA,167,201;%%
%%%%%%%%%%%%%%%%%%%%%%%%%%%%%%%%%%%%%%%%%%%%%%  
 %\cite{Rasheed:1995zv}
\bibitem{Rasheed:1995zv}
  D.~Rasheed,
  %``The Rotating dyonic black holes of Kaluza-Klein theory,''
  Nucl.\ Phys.\  B {\bf 454}, 379 (1995)
  [arXiv:hep-th/9505038].
  %%CITATION = NUPHA,B454,379;%%
%%%%%%%%%%%%%%%%%%%%%%%%%%%%%%%%%%%%%%%%%%%%%%
  %\cite{Larsen:1999pp}
\bibitem{Larsen:1999pp}
  F.~Larsen,
  %``Rotating Kaluza-Klein black holes,''
  Nucl.\ Phys.\  B {\bf 575}, 211 (2000)
  [arXiv:hep-th/9909102].
  %%CITATION = NUPHA,B575,211;%%  
%%%%%%%%%%%%%%%%%%%%%%%%%%%%%%%%%%%%%%%%%%%%%%
%\cite{Sorkin:1983ns}
\bibitem{Sorkin:1983ns}
  R.~D.~Sorkin,
  %``Kaluza-Klein Monopole,''
  Phys.\ Rev.\ Lett.\  {\bf 51}, 87 (1983).
  %%CITATION = PRLTA,51,87;%%
%%%%%%%%%%%%%%%%%%%%%%%%%%%%%%%%%%%%%%%%%%%%%%
%\cite{Gross:1983hb}
\bibitem{Gross:1983hb}
  D.~J.~Gross and M.~J.~Perry,
 % ``Magnetic Monopoles In Kaluza-Klein Theories,''
  Nucl.\ Phys.\  B {\bf 226}, 29 (1983).
  %%CITATION = NUPHA,B226,29;%%
%%%%%%%%%%%%%%%%%%%%%%%%%%%%%%%%%%%%%%%%%%%%%%  
  %\cite{Ishihara:2005dp}
\bibitem{Ishihara:2005dp}
  H.~Ishihara and K.~Matsuno,
 % ``Kaluza-Klein black holes with squashed horizons,''
  Prog.\ Theor.\ Phys.\  {\bf 116}, 417 (2006)
  [arXiv:hep-th/0510094].
  %%CITATION = PTPKA,116,417;%%
%%%%%%%%%%%%%%%%%%%%%%%%%%%%%%%%%%%%%%%%%%%%%%
  %\cite{Brihaye:2006ws}
\bibitem{Brihaye:2006ws}
  Y.~Brihaye and E.~Radu,
%  ``Kaluza-Klein black holes with squashed horizons and d = 4 superposed
%  monopoles,''
  Phys.\ Lett.\  B {\bf 641}, 212 (2006)
  [arXiv:hep-th/0606228].
  %%CITATION = PHLTA,B641,212;%%
%%%%%%%%%%%%%%%%%%%%%%%%%%%%%%%%%%%%%%%%%%%%%%
  %\cite{Wang:2006nw}
\bibitem{Wang:2006nw}
  T.~Wang,
 % ``A rotating Kaluza-Klein black hole with squashed horizons,''
  Nucl.\ Phys.\  B {\bf 756}, 86 (2006)
  [arXiv:hep-th/0605048].
  %%CITATION = NUPHA,B756,86;%%
%%%%%%%%%%%%%%%%%%%%%%%%%%%%%%%%%%%%%%%%%%%%%%
%\cite{Nakagawa:2008rm}
\bibitem{Nakagawa:2008rm}
  T.~Nakagawa, H.~Ishihara, K.~Matsuno and S.~Tomizawa,
%  ``Charged Rotating Kaluza-Klein Black Holes in Five Dimensions,''
  Phys.\ Rev.\  D {\bf 77}, 044040 (2008)
  [arXiv:0801.0164 [hep-th]].
  %%CITATION = PHRVA,D77,044040;%%
%%%%%%%%%%%%%%%%%%%%%%%%%%%%%%%%%%%%%%%%%%%%%%
 %\cite{Tomizawa:2008hw}
\bibitem{Tomizawa:2008hw}
  S.~Tomizawa, H.~Ishihara, K.~Matsuno and T.~Nakagawa,
  %``Squashed Kerr-Godel Black Holes - Kaluza-Klein Black Holes with Rotations
  %of Black Hole and Universe -,''
  Prog.\ Theor.\ Phys.\  {\bf 121}, 823 (2009)
  [arXiv:0803.3873 [hep-th]].
  %%CITATION = PTPKA,121,823;%%
%%%%%%%%%%%%%%%%%%%%%%%%%%%%%%%%%%%%%%%%%%%%%%
%\cite{Matsuno:2008fn}
\bibitem{Matsuno:2008fn}
  K.~Matsuno, H.~Ishihara, T.~Nakagawa and S.~Tomizawa,
  %``Rotating Kaluza-Klein Multi-Black Holes with Godel Parameter,''
  Phys.\ Rev.\  D {\bf 78}, 064016 (2008)
  [arXiv:0806.3316 [hep-th]].
  %%CITATION = PHRVA,D78,064016;%%
%%%%%%%%%%%%%%%%%%%%%%%%%%%%%%%%%%%%%%%%%%%%%%
%\cite{Tomizawa:2008rh}
\bibitem{Tomizawa:2008rh}
  S.~Tomizawa and A.~Ishibashi,
  %``Charged Black Holes in a Rotating Gross-Perry-Sorkin Monopole Background,''
  Class.\ Quant.\ Grav.\  {\bf 25}, 245007 (2008)
  [arXiv:0807.1564 [hep-th]].
  %%CITATION = CQGRD,25,245007;%%
%%%%%%%%%%%%%%%%%%%%%%%%%%%%%%%%%%%%%%%%%%%%%%
  %\cite{Stelea:2008tt}
\bibitem{Stelea:2008tt}
  C.~Stelea, K.~Schleich and D.~Witt,
 %``On squashed black holes in Godel universes,''
  Phys.\ Rev.\  D {\bf 78}, 124006 (2008)
  [arXiv:0807.4338 [hep-th]].
  %%CITATION = PHRVA,D78,124006;%%
 %%%%%%%%%%%%%%%%%%%%%%%%%%%%%%%%%%%%%%%%%%%%%% 
%\cite{Murata:2009jt}
\bibitem{Murata:2009jt} 
  K.~Murata, T.~Nishioka and N.~Tanahashi,
  %``Warped AdS(5) Black Holes and Dual CFTs,''
  Prog.\ Theor.\ Phys.\  {\bf 121}, 941 (2009)
  [arXiv:0901.2574 [hep-th]].
  %%CITATION = ARXIV:0901.2574;%%
 
 %%%%%%%%%%%%%%%%%%%%%%%%%%%%%%%%%%%%%%%%%%%%%%  
  %\cite{Brihaye:2009dm}
\bibitem{Brihaye:2009dm}
  Y.~Brihaye, J.~Kunz and E.~Radu,
  %``From black strings to black holes: Nuttier and squashed AdS(5) solutions,''
  JHEP {\bf 0908}, 025 (2009)
  [arXiv:0904.1566 [gr-qc]].
  %%CITATION = JHEPA,0908,025;%%
  
  %\cite{Tatsuoka:2011tx}
\bibitem{Tatsuoka:2011tx} 
  T.~Tatsuoka, H.~Ishihara, M.~Kimura and K.~Matsuno,
  %``Extremal Charged Black Holes with a Twisted Extra Dimension,''
  arXiv:1110.6731 [hep-th].
  %%CITATION = ARXIV:1110.6731;%%
%%%%%%%%%%%%%%%%%%%%%%%%%%%%%%%%%%%%%%%%%%%%%%  
%\cite{Hoxha:2000jf}
\bibitem{Hoxha:2000jf} 
  P.~Hoxha, R.~R.~Martinez-Acosta and C.~N.~Pope,
  %``Kaluza-Klein consistency, Killing vectors, and Kahler spaces,''
  Class.\ Quant.\ Grav.\  {\bf 17}, 4207 (2000)
  [hep-th/0005172].
  %%CITATION = HEP-TH/0005172;%%
  %%%%%%%%%%%%%%%%%%%%%%%%%%%%%%%%%%%%%%%%%%%%%% 
%\cite{Copsey:2006br}
\bibitem{Copsey:2006br} 
  K.~Copsey and G.~T.~Horowitz,
  %``Gravity dual of gauge theory on S**2 x S**1 x R,''
  JHEP {\bf 0606}, 021 (2006)
  [hep-th/0602003].
  %%CITATION = HEP-TH/0602003;%%
  %%%%%%%%%%%%%%%%%%%%%%%%%%%%%%%%%%%%%%%%%%%%%%
%\cite{Mann:2006yi}
\bibitem{Mann:2006yi} 
  R.~B.~Mann, E.~Radu and C.~Stelea,
  %``Black string solutions with negative cosmological constant,''
  JHEP {\bf 0609}, 073 (2006)
  [hep-th/0604205].
  %%CITATION = HEP-TH/0604205;%%
 %%%%%%%%%%%%%%%%%%%%%%%%%%%%%%%%%%%%%%%%%%%%%%  
  %\cite{Brihaye:2007vm}
\bibitem{Brihaye:2007vm}
  Y.~Brihaye, E.~Radu and C.~Stelea,
  %``Black strings with negative cosmological constant: Inclusion of electric
  %charge and rotation,''
  Class.\ Quant.\ Grav.\  {\bf 24}, 4839 (2007)
  [arXiv:hep-th/0703046].
  %%CITATION = CQGRD,24,4839;%%  
%%%%%%%%%%%%%%%%%%%%%%%%%%%%%%%%%%%%%%%%%%%%%% 
%\cite{Balasubramanian:1999re}
\bibitem{Balasubramanian:1999re} 
  V.~Balasubramanian and P.~Kraus,
  %``A Stress tensor for Anti-de Sitter gravity,''
  Commun.\ Math.\ Phys.\  {\bf 208}, 413 (1999)
  [hep-th/9902121].
  %%CITATION = HEP-TH/9902121;%%
  %%%%%%%%%%%%%%%%%%%%%%%%%%%%%%%%%%%%%%%%%%%%%%
%\cite{Das:2000cu}
\bibitem{Das:2000cu} 
  S.~Das and R.~B.~Mann,
  %``Conserved quantities in Kerr-anti-de Sitter space-times in various dimensions,''
  JHEP {\bf 0008}, 033 (2000)
  [hep-th/0008028].
  %%CITATION = HEP-TH/0008028;%%
  %%%%%%%%%%%%%%%%%%%%%%%%%%%%%%%%%%%%%%%%%%%%%%
%\cite{Skenderis:2000in}
\bibitem{Skenderis:2000in} 
  K.~Skenderis,
  %``Asymptotically Anti-de Sitter space-times and their stress energy tensor,''
  Int.\ J.\ Mod.\ Phys.\ A {\bf 16}, 740 (2001)
  [hep-th/0010138];
  %%CITATION = HEP-TH/0010138;%%
  %%CITATION = HEP-TH 0010138;%%
\\
%\cite{Henningson:1998gx}
%\bibitem{Henningson:1998gx} 
  M.~Henningson and K.~Skenderis,
  %``The Holographic Weyl anomaly,''
  JHEP {\bf 9807}, 023 (1998)
  [hep-th/9806087]; \
  %%CITATION = HEP-TH 9806087;%%
  \\
  M.~Henningson and K.~Skenderis,
  %``Holography and the Weyl anomaly,''
  Fortsch.\ Phys.\  {\bf 48}, 125 (2000)
  [arXiv:hep-th/9812032].
  %%CITATION = HEP-TH 9812032;%%
 %%%%%%%%%%%%%%%%%%%%%%%%%%%%%%%%%%%%%%%%%%%%%% 
  \bibitem{COLSYS}
 U. Ascher, J. Christiansen, R.~D. Russell,
% A collocation solver for mixed order systems of boundary value problems,
 Math. of Comp. {\bf 33} (1979) 659;
 \\
 U. Ascher, J. Christiansen, R.~D. Russell,
%Collocation software for boundary-value ODEs,
 ACM Trans. {\bf 7} (1981) 209.
 %%%%%%%%%%%%%%%%%%%%%%%%%%%%%%%%%%%%%%%%%%%%%% 
  %\cite{McInnes:2010ti}
\bibitem{McInnes:2010ti}
  B.~McInnes,
  %``Fragile Black Holes,''
  Nucl.\ Phys.\  {\bf B842}, 86-106 (2011) 
  [arXiv:1008.0231 [hep-th]].
 %%%%%%%%%%%%%%%%%%%%%%%%%%%%%%%%%%%%%%%%%%%%%% 
 %\cite{McInnes:2011eg}
\bibitem{McInnes:2011eg}
  B.~McInnes,
  %``Kerr Black Holes Are Not Fragile,''
  arXiv:1108.6234 [hep-th].
  %%CITATION = ARXIV:1108.6234;%% 
 %%%%%%%%%%%%%%%%%%%%%%%%%%%%%%%%%%%%%%%%%%%%%% 
  %\cite{Seiberg:1999xz}
\bibitem{Seiberg:1999xz}
  N.~Seiberg and E.~Witten,
  %``The D1 / D5 system and singular CFT,''
  JHEP {\bf 9904}, 017 (1999)
  [arXiv:hep-th/9903224].
  %%CITATION = JHEPA,9904,017;%%


%%%%%%%%%%%%%%%%%%%%%%%%%%%%%%%%%%%%%%%%%%%%%%%%%%%%%%%%%%%%%%%%%%%%%%%%%%%%%%
\end{thebibliography}
\end{document}